\documentclass[english]{llncs}
\usepackage{amsmath,amssymb,verbatim,babel,wrapfig}
\usepackage{graphicx}

\def\FD{Fr\'{e}chet distance}
\def\FSD{free space diagram}

\def\DFSD{double free space diagram}

\def\UD{untangleability space}
\def\UDs{untangleability spaces}
\def\PD{propagation space}

\def\eps{\varepsilon}
\def\simp{diagonal monotonicity test}
\def\fsurface{folded polygon}
\def\fsurfaces{folded polygons}
\def\FSP{Fr\'{e}chet shortest path}
\def\FSPs{Fr\'{e}chet shortest paths}
\def\dcmapping{monotone diagonal mapping}
\def\dcmappings{monotone diagonal mappings}

\begin{document}

\title{Computing the Fr\'{e}chet Distance Between Folded Polygons$^{1}$}

\vspace{-3 mm}

\author{Atlas F. Cook IV$^{2}$, Anne Driemel$^{3}$, Sariel Har-Peled$^{4}$, Jessica Sherette$^{5}$, Carola Wenk$^{5}$}

\institute{}

\footnotetext[1]{\scriptsize This work has been supported by the National Science Foundation grant NSF CAREER \\ CCF-0643597.}
\footnotetext[2]{\scriptsize Department of Mathematics and Computer Science; Eindhoven University of Technology; The Netherlands; a.f.cook@tue.nl}
\footnotetext[3]{\scriptsize Department of Information and Computing Sciences; Utrecht University; The Netherlands; driemel@cs.uu.nl}
\footnotetext[4]{\scriptsize Department of Computer Science; University of Illinois; 201 N. Goodwin Avenue; Urbana, IL, 61801, USA; sariel@uiuc.edu} 
\footnotetext[5]{\scriptsize Department of Computer Science; University of Texas at San Antonio; \\ \{jsherett, carola\}@cs.utsa.edu}

\maketitle

\vspace{-4 mm}

\begin{abstract}
We present the first results showing that the \FD\ between non-flat surfaces can be approximated within a constant factor in polynomial time.  Computing the \FD\ for surfaces is a surprisingly hard problem.  It is not known whether it is computable, it has been shown to be NP-hard, and the only known algorithm computes the \FD\ for flat surfaces (Buchin et al$.$).  We adapt this algorithm to create one for computing the \FD\ for a class of surfaces which we call \fsurfaces.  Unfortunately, if extended directly the original algorithm no longer guarantees that a homeomorphism exists between the surfaces.  We present three different methods to address this problem.  The first of which is a fixed-parameter tractable algorithm.  The second is a polynomial-time approximation algorithm which approximates the optimum mapping.  Finally, we present a restricted class of \fsurfaces\ for which we can compute the \FD\ in polynomial time.
\end{abstract}

\vspace{-2 mm}


\vspace{-4 mm}

\section{Introduction\label{sec:Introduction}}

The \emph{\FD} is a similarity metric for continuous shapes such as curves and surfaces.  In the case of computing it between two (directed open) curves there is an intuitive explanation of the \FD.  Suppose a man walks along one curve, a dog walks along the other, and they are connected by a leash.  They can vary their relative speeds but cannot move backwards.  Such a walk pairs every point on one curve to one and only one point on the other curve (i.e., creates homeomorphism between surfaces) in a continuous way.  The \FD\ of the curves is the minimum leash length required for the man and dog to walk along these curves.  Although less intuitive, the idea is similar for surfaces.

While the \FD\ between polygonal curves can be computed in polynomial time \cite{Alt1995}, computing it between surfaces is much harder.  In \cite{godau} it was shown that even computing the \FD\ between a triangle and a self-intersecting surface is NP-hard.  This result was extended in \cite{bbs-fds-10} to show that computing the \FD\ between 2d terrains as well as between polygons with holes is also NP-hard.  Furthermore, while in \cite{Alt2010} it was shown to be upper semi-computable, it remains an open question whether the \FD\ between general surfaces is even computable.

On the other hand, in \cite{Buchin2006} a polynomial time algorithm is given for computing the \FD\ between two (flat) simple polygons.  This was the first paper to give any algorithm for computing the \FD\ for a nontrivial class of surfaces and remains the only known approach for computing it.  The main idea of their algorithm is to restrict the kinds of different mappings that need to be considered.  Our contribution is to generalize their algorithm to a class of non-flat surfaces we refer to as \fsurfaces.  Given that theirs is the only known approach it is of particular importance to explore extending it to new classes of surfaces.  The major problem we encountered in generalizing the work of \cite{Buchin2006} was that the kinds of different of mappings which need to be considered is less restricted.  We address three different methods to resolve this problem.  In Section \ref{sec:FPTAlgorithm}, we outline a fixed-parameter tractable algorithm.  In Section \ref{sec:GeneralApprox}, we describe a polynomial-time approximation algorithm to compute the \FD\ between \fsurfaces\ within a constant factor.  In Section \ref{sec:LinfParallel}, we describe a nontrivial class of \fsurfaces\ for which the original algorithm presented in \cite{Buchin2006} will compute an exact result.

\vspace{-3 mm}

\section{Preliminaries\label{sec:Preliminaries}}

The \FD\ is defined for two k-dimensional hypersurfaces $P,Q:[0,1]^{x}\rightarrow\mathbb{R}^{d}$, where $x \leq d$,
as \[
\delta_{F}(P,Q)\ =\ \inf_{\sigma:A\rightarrow B}\ \sup_{p\in A}\ \|P(p)-Q(\sigma(p))\|\]
 where $\sigma$ ranges over orientation-preserving homeomorphisms
that map each point $p\in P$ to an image point $q=\sigma(p)\in Q$.  Lastly, $\|\cdot\|$ is the Euclidean norm but other metrics could be used instead.

Let $P,Q:[0,1]^{2}\rightarrow\mathbb{R}^{d}$ be connected polyhedral surfaces for each of which we have a convex subdivision.  We assume that the dual graphs of the convex subdivisions are acyclic, which means that the subdivisions do not have any interior vertices.  We will refer to surfaces of this type as \emph{\fsurfaces}.  We refer to the interior convex subdivision edges of $P$ and $Q$ as \emph{diagonals} and \emph{edges} respectively.  Let $m$ and $n$ be the complexities of $P$ and $Q$ respectively.  Let $k$ and $l$ be the number of diagonals and edges respectively.  Assume without loss of generality the number of diagonals is smaller than the number of edges.  Let $T_{matrixmult}(N)$ denote the time to multiply two $N\times N$ matrices.

\vspace{-3 mm}

\subsection{Simple Polygons Algorithm Summary \label{sec:SimpleSummary}}

In previous work Buchin et al$.$ \cite{Buchin2006} compute the
\FD\ between simple polygons $P$, $Q$.  The authors show that, while the \FD\ between a convex polygon and a simple polygon is the \FD\ of their boundaries, this is not the case for two simple polygons.  The idea of their algorithm is to find a convex subdivision of $P$ and map each of the convex regions of it continuously to distinct parts of $Q$ such that taken together the images account for all of $Q$.

First, they show that the decision
problem $\delta_{F}(P,Q)\leq\eps$ can be solved by (1) mapping
the boundary of $P$, which we denote by $\partial P$, onto the 
boundary of $Q$, which we denote by $\partial Q$, such that $\delta_{F}(\partial P,\partial Q)\leq\eps$
and (2) mapping each diagonal $d$ in the convex subdivision of $P$
to a shortest path $f \subseteq Q$ such that both endpoints of $f$ lie
on $\partial Q$ and such that $\delta_{F}(d,f)\leq\eps$. 

\begin{figure*}[ht]
\centering \includegraphics[width=0.85\textwidth]{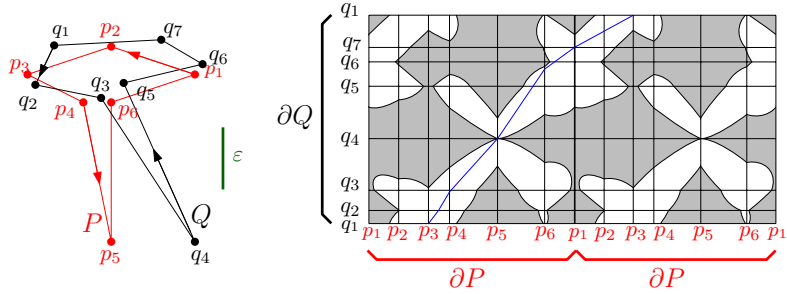}
\centering \caption{\small  The white areas are those in \FSD.  The surfaces are within \FD\ $\varepsilon$ since there is a monotone path starting at the bottom of the \FSD\ and ending at the top which maps every point on the boundary of $P$ exactly once.  This figure was generated using an ipelet created by G\"{u}nter Rote.
         \label{fig:FSDpathNew}}

\vspace{-3 mm}

\end{figure*}

In order to solve subproblem (1) they use the notion of a \FSD.  For open curves $f,g:[0,1]\rightarrow\mathbb{R}^{d}$ it is defined as $FS_{\varepsilon}(f,g) = \{(x,y)\ |\ x \in f,\ y \in g,\ ||x-y||\leq\eps\}$ where $\varepsilon\ \geq 0$.  A monotone path starting at the bottom left corner of the \FSD\ going to the top right exists if and only if the curves are within \FD\ $\varepsilon$.  As shown in \cite{Alt1995}, this can be extended to closed curves by concatenating two copies of the \FSD\ to create a \emph{\DFSD}\ and searching for a monotone path which covers every point in $P$ exactly once see Figure \ref{fig:FSDpathNew}.  
This algorithm can be used to show whether $\delta_{F}(\partial P,\partial Q)\leq\eps$ and find the particular mapping(s) between $\partial P$ and $\partial Q$.   In turn this defines a \emph{placement} of the diagonals, i.e., a mapping of the endpoints of the diagonals to endpoints of the corresponding image curves in $Q$.

Subproblem (2) is solved by only considering paths through the \FSD\ that map a diagonal $d$ onto an image curve $f$ such that $\delta_{F}(d,f)\leq\eps$.  Naturally the particular placement of the diagonals determined in subproblem (1) could affect whether this is true.  Therefore, they must check this for many paths in the \FSD.  Fortunately they can show that it is sufficient to only consider mapping a diagonal to an image curve which is the shortest path between the end points determined by the placement.

Solving these subproblems generates a mapping between $P$ and $Q$ for $\varepsilon$.  This mapping might not be a homeomorphism but the authors show by making very small perturbations of image curves the mapping can be made into one.  These perturbations can be arbitrarily small.  Thus, because the \FD\ is the infimum of all homeomorphisms on the surfaces, the \FD\ is $\varepsilon$.  For simplicity we will refer to these generated mappings as homeomorphisms.  By performing a binary search on a set of critical values they can use the above algorithm for the decision problem to compute the \FD\ of $P$ and $Q$.

\vspace{-3 mm}

\subsection{Shortest Path Edge Sequences \label{sec:ShortestPathEdgeSequences}}

Our algorithm extends the simple polygons algorithm to one for \fsurfaces.  The idea of the algorithm is to subdivide one surface, $P$, into convex regions and pair those with corresponding regions in the other surface, $Q$.  The difference is that those regions of $Q$ are now \fsurfaces\ rather than just simple polygons.  The authors of \cite{Buchin2006} show that the \FD\ of a convex polygon and a simple polygon is just the \FD\ of their boundaries.  Using almost the same argument we prove that it also holds for \fsurfaces, see Appendix \ref{sec:ConvexVSFoldedPolygonProof} for the full proof 

\begin{lemma}
\label{lem:GeneralApprox-ConvexVSFoldedPolygon}
The \FD\ between a \fsurface\ $P$ and a convex polygon $Q$ is the same as that between their boundary curves.
\end{lemma}

\begin{figure*}[ht]

\vspace{-4 mm}

\centering \includegraphics[width=0.80\textwidth]{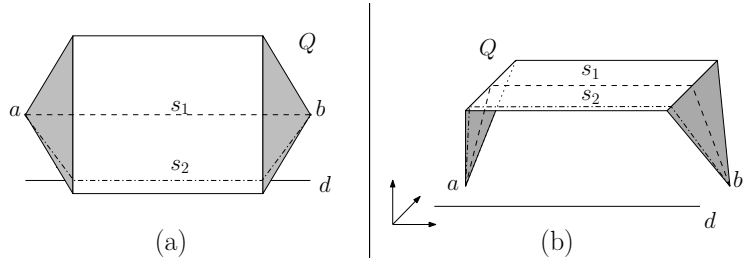}
\centering \caption{\small An example where the image curve $s_2$ with the smallest \FD\ to a diagonal $d$ is a non-shortest path $s_1$ in $Q$. (a) overhead view, (b) sideview.
         \label{fig:SPCounterExample}\label{fig:FSDpath}}

\vspace{-3 mm}

\end{figure*}

As mentioned in Section \ref{sec:SimpleSummary}, for the simple polygon algorithm it suffices to map diagonals onto shortest paths between two points on $\partial Q$. By contrast, there are \fsurfaces\ where a homeomorphism between the surfaces does not exist when diagonals are mapped to shortest paths but does exist when the paths are not restricted, see Figure \ref{fig:FSDpath}.  The curve $s_1$ is the shortest path between the points $a$ and $b$ but the curve $s_2$ has smaller \FD\ to $d$ than $s_1$ has.  We must therefore consider mapping the diagonals to more general paths.  Fortunately, we can show that these more general paths still have some nice properties for \fsurfaces.

\vspace{-3 mm}

\begin{lemma}
\label{lem:EdgePath}
Let $Q$ be a \fsurface, $u$ and $v$ be points such that $u,v \in \partial Q$, $E = \lbrace e_0, e_1, \ldots, e_s \rbrace$ be a sequence of edges in the convex subdivision of $Q$, and $d$ be a line segment.  Given $\varepsilon > 0$ we can find a curve $f$ in $Q$ that follows the edge sequence $E$ from $u$ to $v$ such that $\delta_F (d,f) \leq \varepsilon$, if such a curve exists, in $O(s)$ time.
\end{lemma}

\proof
We construct a series $F = FS_{\varepsilon}(e_0,d), FS_{\varepsilon}(e_1,d), \ldots, FS_{\varepsilon}(e_s,d)$, of 2-dimensional free space diagrams.  Any two edges $e_i$ and $e_{i+1}$ are on the boundary of the same convex polygon in the convex subdivision of $Q$ so we can assume without loss of generality that $f$ consists of straight line segments between the edge intersections.  This is similar to the shortcutting argument used to prove Lemma 3 in \cite{Buchin2006}.  Thus, we only need to check the points where $f$ crosses an edge of $Q$.  For $\delta_F (d,f) \leq \varepsilon$ to be true, the preimages of those crossing points must be monotone along $d$.  Let $FS_{\varepsilon}^{'}(e_i,d)$ be the projection of $FS_{\varepsilon}(e_i,d)$ onto $d$.  Let $F^{'} = FS_{\varepsilon}^{'}(e_1,d), FS_{\varepsilon}^{'}(e_2,d), \ldots, FS_{\varepsilon}^{'}(e_s,d)$.

To verify that the preimage points on $d$ can be chosen such that they are monotone, we check the intervals of $F^{'}$.  Specifically, for $i<j$, the point on $d$ mapped to $e_i$ must come before the one mapped to $e_j$.  This can be checked by greedily scanning left to right and always choosing the smallest point on $d$ which can be mapped to some edge.  A search of this form takes $O(s)$ time.
\qed

The dual graph of the faces of $Q$ is acyclic.  This implies that there is a unique sequence of faces through which a shortest path from $u$ to $v$, where $u,v \in \partial Q$, must pass.  Necessarily, there must also be a unique edge sequence that the shortest path follows.  We call such an edge sequence the \emph{shortest path edge sequence}.

\begin{lemma}
\label{lem:OnlyShortestPathEdgeSequenceNeeded}
Let $d$ be a diagonal.  If there is a curve $f \subseteq Q$ with $\delta_F (d,f) \leq \varepsilon$ then there is a curve $g \subseteq Q$ which follows the \emph{shortest path edge sequence} such that $\delta_F (d,g) \leq \varepsilon$.
\end{lemma}

\proof
Let $E_f$ and $E_g$ be the edge sequences of $f$ and $g$ respectively.  By definition the dual graph of the faces of $Q$ is acyclic, so $E_g$ must be a subsequence of $E_f$.  $E_g$ induces a sequence of free space intervals.  If there is a monotone path in the free space interval sequence induced by $E_f$, we can cut out some intervals and have a monotone path in the free space for $E_g$. 
\qed

From Lemma \ref{lem:OnlyShortestPathEdgeSequenceNeeded}, we just need to consider paths that follow the shortest path edge sequence.  We refer to paths that follow this edge sequence and consist of straight line segments between edges as \emph{\FSPs}.  In addition, $s$ in Lemma \ref{lem:EdgePath} is bounded by the number of edges in along the shortest path edge sequence between $u$ and $v$, and $E$ will be the shortest path sequence.  This implies the following theorem.

\begin{theorem}
\label{thm:FrechetPathWithPoints}
Let $Q$ be a \fsurface, $u$ and $v$ be points such that $u,v \in \partial Q$, and $d$ be a line segment.  Given $\varepsilon > 0$, we can in $O(l)$ time find a curve $f$ in $Q$ from $u$ to $v$ such that $\delta_F (d,f) \leq \varepsilon$ if such a curve exists.
\end{theorem}

Suppose we have a homeomorphism between $\partial P$ and $\partial Q$.  The endpoints of the image curves must appear on $\partial Q$ in the same order as their respective diagonal endpoints on $\partial P$.  The homeomorphism also induces a direction on the diagonals in $P$ and on the edges in $Q$.  Specifically, we consider diagonals and edges to start at their first endpoint along $\partial P$ or $\partial Q$ respectively in a counterclockwise traversal of the boundaries.  We denote by $D_{e}$ the set of diagonals whose associated shortest
path edge sequences contain an edge $e \subseteq Q$. Observe that pairwise
non-crossing image curves must intersect an edge $e$
in the same order as their endpoints occur on $\partial Q$. We refer
to this as the \emph{proper intersection order} for an edge $e$.

\vspace{-3 mm}

\subsection{Diagonal Monotonicity Test and Untangleability \label{sec:DiagonalConsistency}}

We now define a test between two \fsurfaces\ $P$ and $Q$ which we call the \emph{\simp}.  For a given $\varepsilon$ this test returns true if the following two things are true.  First, $\delta_F(\partial P, \partial Q) \leq \varepsilon$.  Second, for every diagonal $d_i$ in the convex subdivision of $P$, the corresponding \FSP\ $f_i$ in $Q$ has $\delta_F(d_i, f_i) \leq \varepsilon$.  We refer to the class of mappings of the \fsurfaces\ generated by this test as \emph{\dcmappings}.  This is similar to the test used by \cite{Buchin2006} except ours uses \FSPs\ instead of the shortest paths.

Unfortunately, because the image curves of the diagonals are no longer shortest paths, they may cross each other and we will no longer be able to generate a mapping between the \fsurfaces\ which is a homeomorphism.  For an example of this see Appendix \ref{sec:CounterExample}.  Thus the \simp\ might return true when in fact a homeomorphism does not exist.  We must explicitly ensure that the image curves of all diagonals are non-crossing.  In particular, we refer to a set of image curves $F = \lbrace f_1 \ldots f_k \rbrace$ as \emph{untangleable} for $\varepsilon$ if and only if there exists a set of image curves $F^{'} = \lbrace f^{'}_1 \ldots f^{'}_k \rbrace$ where $f_i$ and $f^{'}_i$ have the same end points on $\partial Q$, $\delta_F (d_i,f^{'}_i) \leq \varepsilon$, and the curves of $F^{'}$ are pairwise non-crossing.  A homeomorphism exists between the \fsurfaces\ for $\varepsilon$ if and only if there exists a \dcmapping\ whose image curves are untangleable for $\varepsilon$.  This follows from the same argument used in \cite{Buchin2006} for simple polygons.

As shown in Theorem \ref{thm:FrechetPathWithPoints}, computing \FSPs\ instead of shortest paths does not increase the asymptotic run time.  To optimize this $\varepsilon$ we can perform a binary search on a set of critical values.  As in \cite{Buchin2006}, the number of critical values is $O(m^2 n + m n^2 )$.  The three types of critical values between a diagonal and its corresponding path through $Q$ are very similar to those outlined in the simple polygons algorithm.  So, by following the paradigm set forth by \cite{Buchin2006}, we arrive at the following theorem:

\begin{theorem}
\label{thm:SimpOptimized-Runtime}
The minimum $\varepsilon$ for which two \fsurfaces\, $P$ and $Q$, pass the \simp\ can be computed in time $O(kT_{matrixmult}(mn) \log(mn))$.
\end{theorem}

\vspace{-3 mm}

\section{Fixed-Parameter Tractable Algorithm\label{sec:FPTAlgorithm}}

In this section we outline an algorithm to decide for a fixed mapping between the boundaries of a pair of \fsurfaces\ whether the image curves induced from the mapping are untangleable.  From this we create a fixed-parameter tractable algorithm for computing the \FD\ between a pair of \fsurfaces.

\vspace{-3 mm}

\subsection{Untangleability Space\label{sec:UDiagram}}

Let $e$ be an edge in $Q$ which is crossed by the image curves of $h$ diagonals, $d_1, \ldots, d_h$.  We assume without loss of generality that the image curves of the diagonals cross $e$ in proper intersection order if, for all $1 \leq i,j \leq h$ where $i<j$, the image curve of $d_i$ crosses $e$ before the image curve of $d_j$ crosses $e$.  Let the {\em \UD} $U_{e}$ contain all $k$-tuples of points on the diagonals which can be mapped to crossing points on the edge $e$ within distance $\varepsilon$ and such that the crossing points are in the proper intersection order along $e$.  $U_{e}$ can be shown to be convex yielding the following theorem.

\begin{theorem}
\label{thm:ConvexUntangleabilitySpace}
$U_e(d_1, \ldots, d_k)$ is convex.

\end{theorem}

This theorem can be proven by linearly interpolating between points in $U_{e}$.  For the full proof see Appendix \ref{sec:UntangleSpaceConvex}.

\vspace{-3 mm}

\subsection{Fixed-Parameter Tractable Algorithm\label{sec:Algorithm}}
\begin{minipage}{\textwidth}
We assume the complexity of $k$ and $l$, the convex subdivisions of $P$ and $Q$ are constant.  Checking for the existence of a set of image curves which are untangleable can be done by using the \UDs\ of the diagonals.  Assume we are given some homeomorphism between $\partial P$ and $\partial Q$ from which we get a placement of the diagonals.  We first choose an edge in $Q$ to act as the root of the {\em edge tree} that corresponds to the dual graph of $Q$.  We propagate constraints imposed by each \UD\ up the tree to the root node to determine if the set of image curves induced by the placement of the diagonals is untangleable.

\begin{wrapfigure}{r}{0.15\textwidth}
	\centering \includegraphics[width=0.14\textwidth]{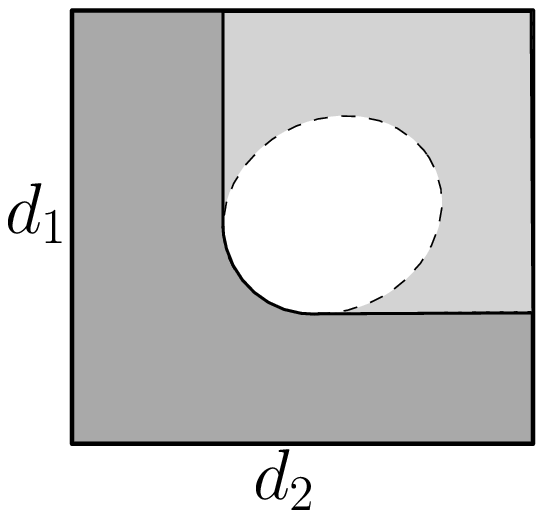}
	\centering \caption{}
	\label{fig:Propogation-01}
\end{wrapfigure}

\hspace{4 mm} The \UD\ of an edge $e$, $U_{e}$, contains exactly those sets of points on the diagonals in $D_e$ which can be mapped to the edge in the proper intersection order.  The point chosen in $U_{e}$ imposes a constraint on what points may be chosen in other \UDs.  In particular the corresponding points on all of the diagonals must be monotone with respect to their edge sequence.  We define $C(U_{e_{i}})$ as the Minkowski sum of $U_{e_{i}}$ with a ray in the opposite direction of the constraint on each of the diagonals in $D_{e_{i}}$, see Figure \ref{fig:Propogation-01}.  In Figure \ref{fig:Propogation-01} $U_e$ is shown in white and $C(U_e)$ is the union of the white and light gray portions.  The direction of this constraint depends on which side of the edge $e_{i}$ the next edge is.  $C(U_{e_{i}})$ contains exactly those sets of points on the diagonals not excluded from having a monotone mapping with $U_{e_{i}}$.

\end{minipage}

We define for every edge $e$ a $k$-dimensional {\em \PD} $P_{e}$. If $e$ is a leaf in the tree, then $P_{e}=U_{e}$.  Otherwise, define\[P_e=U_e \cap C(P_{e_1}) \cap \ldots \cap C(P_{e_j})\] where $e$ is the parent of the edges $e_{1}, e_{2}, \ldots, e_{j}$.  $C(P_{e_{j}})$ contains only those points that are not excluded by the constraints of the tree rooted at $e_j$ from being used to untangle on the parent of $e_{j}$. The \PD\ for the root will be empty if and only if this set of image curves are not untangleable.  From our assumptions, the \PD\ of the root can be computed in constant time as the intersection of semi-algebraic sets \cite{Basu2006}.  Let $F(k,l)$ be the time complexity of computing this intersection.  

Consider two different mappings between $\partial P$ and $\partial Q$.  These determine different placements of the diagonals.  If all of the image curves of all of the diagonals have the same shortest path edge sequence in both of the mappings the test will return the same result.  Thus, we only need to test paths through the \FSD\ which cross the diagonals and edges in a different order.  The \FSD\ for $\partial P$ and $\partial Q$ contains $2k$ vertical line segments that will each contribute $O(kl)$ different mappings of the diagonals and edges. Hence, there are $O((kl)^{2k})$ paths through the \FSD\ which we need to test.  For each of these we can check whether a global untangling exists as described above in constant time.  Similar to the algorithm for polygonal curves \cite{Alt1995} we can perform Cole's \cite{Cole1987} technique for parametric search \cite{Megiddo1983} to optimize the value of $\varepsilon$.  For convex subdivisions with constant number of edges $l$ and diagonals $k$, this yields a fixed-parameter tractable algorithm with runtime polynomial in $m$ and $n$.

\begin{theorem}
\label{thm:FixedParameter-Algorithm}
We can compute the \FD\ of two \fsurfaces\ in time $O((F(k,l)(kl)^{2k} + kT_{matrixmult}(mn))\log(mn))$.
\end{theorem}

\vspace{-3 mm}

\section{Constant Factor Approximation Algorithm\label{sec:GeneralApprox}}

In this section we present an approximation algorithm based on our \simp\ to avoid tangles altogether.  First we demonstrate the following theorem:

\begin{theorem}
\label{thm:GeneralApprox-SimpleTest}
If two \fsurfaces, $P$ and $Q$, pass the \simp\ for some $\varepsilon$, then $\delta_F(P, Q) \leq 9\varepsilon$.
\end{theorem}

\proof
Consider the image curves of the diagonals of $P$ found by performing our \simp.  To pass the \simp\ there must be a homeomorphism between $\partial P$ and $\partial Q$.  The image curves of the diagonals will be mapped in the proper order along the boundary of $Q$.  Therefore, if a pair of image curves cross in $Q$ they must do so an even number of times.

Take two consecutive points $u$, $v$ on the boundary of $P$ that are connected by a diagonal $d= \overline{uv}$, and consider the convex ear of $P$ that $d$ ``cuts off''.  Let $d^{'}$ be the image curve of $d$ in $Q$.  In order to create a homeomorphism between $P$ and $Q$, $d^{'}$ should cut off an ear of $Q$ which can be mapped to the ear of $P$.  Unfortunately, some image curves may cross this $d^{'}$ and cause tangles.  Consider the arrangement of image curves in Q and let $d^{''}$ be the highest level of this arrangement closest to the top of the ear ($\partial Q$), such that $d^{''}$ connects the image points $u$ and $v$ on the boundary of $Q$.

\begin{figure}[ht]

\vspace{-6 mm}

     \centering \includegraphics[width=0.9\textwidth]{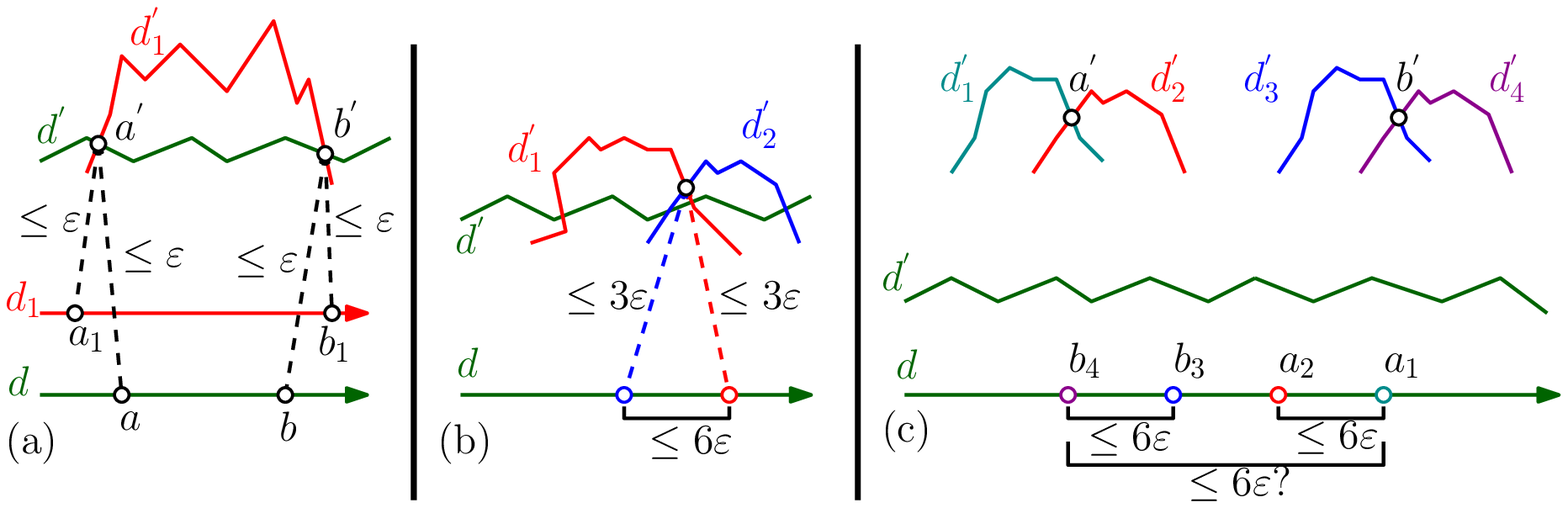}
     \centering \caption{\small
     (a) the intersections between $d$ and $d_1$ imply that $\overline{ab}$ can be mapped to the region of $d_1^{'}$ between $a^{'}$ and 
$b^{'}$ within \FD\ $\varepsilon$
     (b) this is the preimage of the intersection between $d_1^{'}$ and $d_2^{'}$ on $d$,
     (c) an example of the preimages of two intersections occurring out of order of $d$}
     \label{fig:FoldedApprox-02}

\vspace{-4 mm}

\end{figure}

Observe that if an image curve $d_1^{'}$ crosses $d^{'}$ from below then that intersection point $a^{'}$ has a pre-image on both $d$ and $d_1$.  These points $a$ on $d$ and $a_1$ on $d_1$ can be no more than $2\varepsilon$ apart since they both map to the intersection within distance epsilon.  In addition, $d_1^{'}$ must cross back below $d^{'}$ eventually since all image curves which cross do so an even number of times (take the first such occurrence after the initial crossing).  The preimage points $b$ and $b_1$ of this second intersection $b$ are also no more than $2\varepsilon$ apart.  Since both $d$ and $d^{'}$ are line segments, every point on the line segment $\overline{ab}$ on $d$ is $\leq 2 \varepsilon$ distance from some point on the line segment $\overline{a_{1}b_{1}}$ on $d_1$.  For an approximation of $3\varepsilon$ we can map a point on $d$ to its corresponding point on $d_1$ and then to where that point maps on $d_1^{'}$, see Figure \ref{fig:FoldedApprox-02}(a).  Hence, the diagonal $d$ can be mapped to an image curve on or above $d_1^{'}$ within \FD\ $3\varepsilon$.

If this image curve $d_1^{'}$ then crosses another image curve $d_2^{'}$ this argument above cannot be just repeated because the approximation factor would depend linearly on the number of image curves which cross each other.   The preimage points on $d$ of such an intersection not involving $d^{'}$ are separated by at most $6\varepsilon$.  This is because both diagonals involved in the intersection have a $3\varepsilon$ correspondence between the region of them mapped above $d^{'}$ and $d$.  If the preimages are in order there is no problem.  If they occur out of order they cause a monotonicity constraint.  Fortunately, we can collapse this region on $d$ to the leftmost preimage with $6\varepsilon$ and then map it to the corresponding point on $d_2^{'}$ in $3\varepsilon$ for a total of $9\varepsilon$, see Figure \ref{fig:FoldedApprox-02}(b).  If the preimage points in this example were reversed they would be in order.

Thus, we can approximate away the monotonicity constraint of single intersections with $9\varepsilon$.  We must also verify that if the preimages of single intersections occur out of order it does not effect our approximation, see Figure \ref{fig:FoldedApprox-02}(c).  Due to lack of space the discussion of these technical cases has been moved to the appendix, see Appendix \ref{sec:CFATechnicalCases}.

From these we get that $\delta_F(d,d^{''}) \leq 9\varepsilon$.  Now collapse the ear we initially selected in $P$ to $d$.  Likewise in $Q$ collapse the corresponding ear to $d^{''}$.   This is okay because $d^{''}$ is above all of the other image curves in Q.   This pairs the ear we cut off of $P$ with the part of Q above $d^{''}$ which is a \fsurface\ (it could be simpler than a \fsurface\ but we know it's no more complex than that).  By Lemma \ref{lem:GeneralApprox-ConvexVSFoldedPolygon} the ear of $P$ and the \fsurface\ above $d^{''}$ must be within \FD\ $9\varepsilon$.

Choose another ear in $P$.  We can repeat the above arguments to remove this new ear and its corresponding ear in $Q$.  The dual graph of $P$ is a tree.  Each time we repeat this argument we are removing a leaf from the tree.  Eventually, the tree will contain only a single node which corresponds to some triangle in $P$ which we map to the remainder of $Q$.  
\qed

As a direct consequence of Theorems \ref{thm:GeneralApprox-SimpleTest} and \ref{thm:SimpOptimized-Runtime} we get the following theorem:

\begin{theorem}
\label{thm:GeneralApprox-Algorithm}
We can compute a $9$-approximation of the \FD\ of two \fsurfaces\ in time $O(kT_{matrixmult}(mn)\log(mn))$.
\end{theorem}

\vspace{-3 mm}

\section{Axis-Parallel Folds and $L_{\infty}$ Distance \label{sec:LinfParallel}}

The counter example from Appendix \ref{sec:CounterExample} works for all $L_p$ except for $L_{\infty}$.  In this section we outline a special case where using the $L_{\infty}$ metric guarantees that if a pair of \fsurfaces\ pass the \simp\ for $\varepsilon$ their \FD\ is no more than $\varepsilon$.  Specifically, if all of the line segments in the convex subdivision of the surfaces are parallel to the x-axis, y-axis, or z-axis, we show that it is sufficient to use shortest paths instead of \FSPs.  Since shortest paths never cross we can use the simple polygons algorithm in \cite{Buchin2006} to compute the \FD\ of the surfaces.  We first prove the following lemma.

\begin{lemma}
\label{lem:HalfSpaceRestrictions}
Let $R$ be a half-space such that the plane bounding it, $\partial R$, is parallel to the xy-plane, yz-plane, or xz-plane.  Given a \fsurface\ $Q$ with edges parallel or perpendicular to the x-axis and points $a,b \in Q \cap R$, let $f$ be a path in $Q$, which follows the shortest path edge sequence between $a$ and $b$.  If $P$ is completely inside of $R$ so is the shortest path $f^{'}$ between $a$ and $b$.
\end{lemma}

For the lemma to be false there must exist a $Q$, $R$, and $f$ which serve as a counter example.  There must be at least one edge $e_j$ in $Q$  such that $f \cap e_j \in R$ and $f^{'} \cap e_j \not\in R$.  In particular, let $e_j$ be the first edge where this occurs along the shortest path edge sequence.  First consider a $Q$ where all of the edges of it are perpendicular to $\partial R$.  A line segment in the shortest path $f^{'}$ connects the endpoints of two edges in $Q$.  Let $e_i$ and $e_k$ be the edges that define the line segment in $f^{'}$ that passes through $e_{j}$.  We now consider several cases in how those edges are positioned.

\begin{figure}[ht]
     \centering \includegraphics[width=0.90\textwidth]{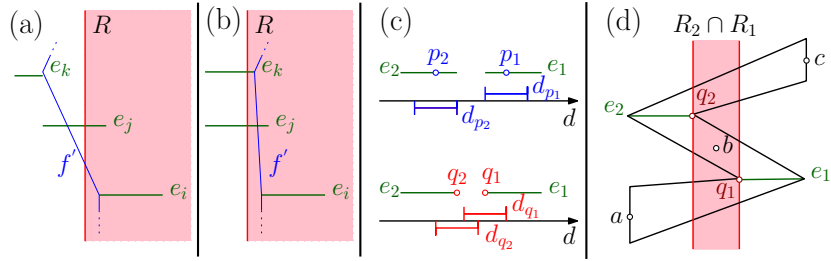}
\centering \caption{\small (a), (b) are examples of case (I) and case (II).  (c) example intervals for the two different paths.  (d) together the edges $e_1$ and $e_2$ cause a monotonicity constraint.}
     \label{fig:ParallelPerpTangleExpNew-02}

\vspace{-4 mm}

\end{figure}

Case (I) occurs when $e_k$ is completely outside of $R$, see Figure \ref{fig:ParallelPerpTangleExpNew-02}(a).  While this does force $f^{'}$ to cross $e_j$ outside of $R$, there is no $f$ which can pass through $e_k$ while remaining inside of $R$.  Because $Q$ is a \fsurface\ any path between $a$ and $b$ must path through the edges in the shortest path edge sequence including $e_k$.  Thus no $f$ can exist entirely within $R$.

Case (II) occurs when part of $e_k$ is in $R$ and $f^{'}$ crosses it in the part in $R$, see Figure \ref{fig:ParallelPerpTangleExpNew-02}(b).  In this case $f^{'}$ does not cross $e_j$ outside of $R$.

Due to space limitations the discussion of the remaining cases has been moved to the appendix, see Appendix \ref{sec:Lemma4Appendix}.  Each of the remaining cases can be reduced to these first two. Using this lemma we can prove the following theorem:

\begin{theorem}
\label{thm:ParallelPerpCase}
The \FD\ between two surfaces, both with only diagonals/edges parallel to the x-axis, y-axis, or z-axis, can be computed in time $O(kT_{matrixmult}(mn)\log(mn))$.
\end{theorem}

\vspace{-3 mm}

\proof Let $Q$ be a \fsurface, $d$ be a diagonal, and $f^{'}$ be the shortest path between points $a$ and $c$ on $\partial Q$.  Using Lemma \ref{lem:HalfSpaceRestrictions} we prove that if there exists a \FSP\ $f$ between points $a$ and $c$ such that $\delta_F(d,f) \leq \varepsilon$, then $\delta_F(d,f^{'}) \leq \varepsilon$.

\vspace{-3 mm}

\subsubsection*{Minkowski Sum Constraints}

Since we are using the $L_{\infty}$ distance, the unit ball is a cube with sides of length 1.  The Minkowski sum of a diagonal $d$ in $P$ and a cube of side length $\varepsilon$ yields a box.  Points in the diagonal $d$ can only map to points in this region.  It can be defined by the intersection of 6 half-spaces; all of these have boundaries parallel to either the xy-axis, the xz-axis, or the yz-axis.  Thus, from Lemma \ref{lem:HalfSpaceRestrictions}, we know that if any path through $Q$ is completely within this box, then the shortest path $f^{'}$ will be, too.  This means that for each edge $e_i$ on the shortest path edge sequence $f^{'} \cap e_i$ is within distance $\varepsilon$ of some non-empty interval of $d$.

\vspace{-3 mm}

\subsubsection*{Monotonicity Constraints}

For the shortest path $f^{'}$ between the boundary points to have $\delta_F(d,f^{'}) > \varepsilon$, at least two of these intervals must be disjoint and occur out of order along $d$, see Figure \ref{fig:ParallelPerpTangleExpNew-02}(c).  Such a case introduces a monotonicity constraint on $\varepsilon$.  If no such intervals existed then we could choose a monotone sequence of points along $d$ such that each point is within distance $\varepsilon$ of an edge and the sequence of edges they map to would have the same order as the shortest path edge sequence showing that $\delta_F(d,f^{'}) \leq \varepsilon$.

Let $e_1$ and $e_2$ be two edges along the shortest path edge sequence for which such bad intervals occur.  Let $p_1$ and $p_2$ be points on the shortest path where it intersect edges $e_1$ and $e_2$ respectively.  Let $q_1$ and $q_2$ be the same for $f$.  Finally, let $d_{r}$ contain all of the points on $d$ which are within distance $\varepsilon$ of the point $r$.  Since $\delta_F(d,f) \leq \varepsilon$, $d_{q_1}$ and $d_{q_2}$ must overlap or occur in order along $d$.

Let $R_1$ be the half-space whose bounding plane contains $q_1$ and is perpendicular to $d$.  likewise let $R_2$ be the half-space whose bounding plane contains $q_2$ and is perpendicular to $d$, see Figure \ref{fig:ParallelPerpTangleExpNew-02}(d).   Let $R_1$ extend to the left along $d$ and $R_2$ extend to the right along $d$.  $\partial R_2$ must occur before $\partial R_1$ along $d$ or the edges are in order and no monotonicity constraint is imposed.  Assume $R_1$ encloses all of $f^{'}$ between $a$ and $e_2$.  If it does not we can choose a new edge between $a$ and $e_2$ to use as $e_1$ for which this is true.  Doing so only increases the monotonicity constraint.  Likewise we can assume $R_2$ encloses all of $f^{'}$ between $e_2$ and $c$.

Assume $a$, $b$, and $c$ lie on $f^{'}$.  Specifically, let $a$ and $c$ be the end points of $f^{'}$ on $\partial Q$.  Naturally, a shortest path must exist between $a$ and $c$ and it must contain at least one point in $R_1 \cap R_2$ which we call $b$.  $f$ follows the shortest path edge sequence between $a$ and $c$, so it must also cross all of the edges in the shortest path edge sequence between $a$ and $b$.  Therefore, to show that $p_2$ is inside of $R_1$ we can directly apply Lemma \ref{lem:HalfSpaceRestrictions} to the points $a$ and $b$.  A similar method can be used for $e_2$ with points $b$ and $c$ to show $p_1$ is inside $R_2$.  Since $d_{q_1}$ and $d_{q_2}$ overlap or are in order, $d_{p_1}$ and $d_{p_2}$ must as well.  Therefore, $\delta_F(d,f^{'}) \leq \varepsilon$ and shortest paths can be used for this variant of \fsurfaces\ instead of \FSPs.  Because we are using shortest paths we can just use the simple polygons algorithm.  This yields Theorem \ref{thm:ParallelPerpCase}. \qed

\vspace{-3 mm}

\section{Future Work\label{sec:Conclusion}}

The constant factor approximation outlined in Section \ref{sec:GeneralApprox} can likely still be improved.  Specifically, we consider only the worst case for each of the out-of-order mappings which may not be geometrically possible to realize.  In addition, we currently approximate the \FD\ by mapping image curves one-by-one to the top of the arrangement of other image curves.  It would of course be more efficient to untangle image curves by mapping them to some middle curve rather than forcing one to map completely above the others.

Finally, while the problem of untangling seems hard, it is also possible that a polynomial-time exact algorithm could exist.  The acyclic nature of our surfaces seems to limit the complexity of our mappings.  The methods used to prove that computing the \FD\ between certain classes of surfaces is NP-hard in \cite{bbs-fds-10} are not easy to apply to \fsurfaces.

\vspace{-4 mm}

\bibliographystyle{lipics}\setlength{\itemsep}{-2mm}
\bibliography{bibtex}

\newpage

\appendix
\section{ Proof of Lemma \ref{lem:GeneralApprox-ConvexVSFoldedPolygon} \label{sec:ConvexVSFoldedPolygonProof}}

\proof
Assume that the \FD\ between the boundary curves is $\varepsilon$.  Now consider mapping the diagonals in the convex decomposition of $P$ to $Q$.  Because $Q$ is convex, we can connect the images of the endpoints of each of the diagonals with a line segment.  These line segments subdivide $Q$ into a set of convex polygons each of which are paired with a convex polygon in $P$.  The boundaries of the paired polygons must be within \FD\ $\varepsilon$ of each other, consequently, by \cite{Buchin2006}, the paired convex polygons must be in \FD\ $\varepsilon$.  

Combining all of these mappings, we have a homeomorphism between $P$ and $Q$.  Thus the \FD\ of $P$ and $Q$ must also be $\leq \varepsilon$.
\qed

\section{Counter Example \label{sec:CounterExample}}

\begin{figure}[ht]
     \centering \includegraphics[width=\textwidth]{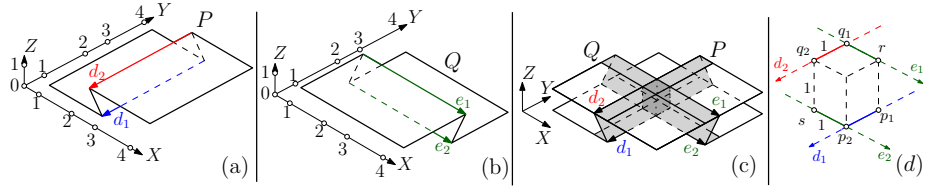}
     \centering \caption{\small (a),(b) The \fsurfaces\ P and Q,
              (c) P and Q overlaid such that the zigzag portions of the surfaces outline a cube at their center,
              (d) The cube has sides of length one.  The marked points must be used in the mappings for $\varepsilon=1$ but they 
                  can be shown to be mutually exclusive}
     \label{fig:PandQ}
\end{figure}

In this section we show a counter example to the claim that it is sufficient that a pair of \fsurfaces\ pass the \simp\ for $\varepsilon$ to guarantee that they are within in \FD\ $\varepsilon$.  Let P and Q be the surfaces shown in Figures \ref{fig:PandQ}(a) and \ref{fig:PandQ}(b) with $\varepsilon = 1$.

The surfaces are the same except rotated by $90$ degrees.  Each surface consists of two parallel 2 by 3 rectangles which are connected together by a smaller slanted rectangle.  One of the large rectangles is at height $1$ while the other is at height $0$.  Figure \ref{fig:PandQ}(c) shows how the two surfaces are overlaid.

\begin{figure}[ht]
     \centering \includegraphics[width=0.95\textwidth]{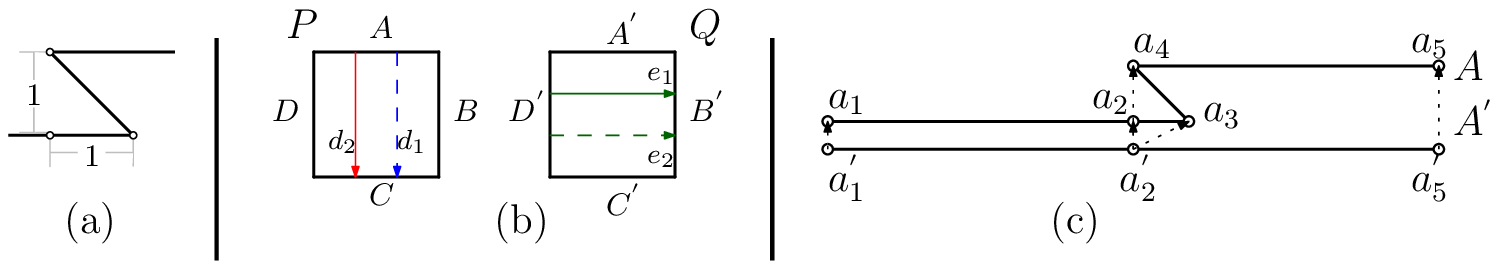}
     \centering \caption{\small (a) Sideview of the surfaces,
              (b) Top view of P and Q showing which sides are mapped to which,
              (c) A side of P is mapped onto a side of Q.  $a_1$ and $a_1^{'}$ as well as $a_2$ and $a_2^{'}$ occupy the same
points in space and a separation is shown for clarity.}
     \label{fig:Sideview}
\end{figure}

We can prove that the boundary curves have \FD\ $1$ to each other.  Pair up each side of P with one in Q as indicated in Figure \ref{fig:Sideview}(b).  These paired portions of the boundary have \FD\ $1$ to each other.  The case of mapping $A$ to $A^{'}$ is given in Figure \ref{fig:Sideview}(c).  For the beginning and ending of the sides the curves either overlap or are exactly distance $1$ apart with one directly above the other, so the mapping used is obvious.  Matching the flat middle part on $A^{'}$ to the zigzag in $A$ is a bit harder.  The key insight is to observe that the point $a_2^{'}$ is within distance $1$ to $a_2$, $a_3$, and $a_4$.  The entire zigzag can be mapped to this point.  The portion after the point $a_4^{'}$ can then be mapped straight down.

Consider the intersection between $P$ and $Q$.  At the center of the \fsurfaces\ is a cube with side length $1$, see Figures \ref{fig:PandQ}(c) and \ref{fig:PandQ}(d).  The point $p_1$ on $d_1$ can map both to the point $p_2$ on edge $e_2$ and to the point $r$ on edge $e_1$ within distance $1$.  Thus, there exists an image curve $f_1$ such that $\delta_F(d_{1},f_{1}) \leq 1$.  Likewise, the point $q_2$ on $d_2$ can map both to the point $s$ on edge $e_2$ and to the point $q_1$ on edge $e_1$ within distance $1$, so there exists an image curve $f_2$ such that $\delta_F(d_{2},f_{2}) \leq 1$.  Therefore, $P$ and $Q$ pass the \simp\ for $1$.

Notice that $p_1$ is the only point on $d_1$ which can map to $e_1$.  Given the direction of $d_1$, $f_1$ starts at the top of $Q$ and must cross edge $e_2$ then $e_1$.  Thus no points after $p_1$ on $d$ can mapped to $e_2$ so $p_2$ is the only reachable point on $e_2$.  Likewise, $s$ is the only point on $e_2$ the diagonal $d_2$ can map to.  $s$ comes before $p_2$ along $e_2$ so the image curve of $d_2$ must cross $e_2$ before the image curve of $d_1$.  This is different from the proper intersection order that comes from the boundary.  Therefore, even though $P$ and $Q$ passed the \simp\ for $1$, $\delta_F(P,Q) > 1$.

\section{Convexity of the Untangleability Space \label{sec:UntangleSpaceConvex}}
\begin{figure}[ht]
     \centering \includegraphics[width=0.80\textwidth]{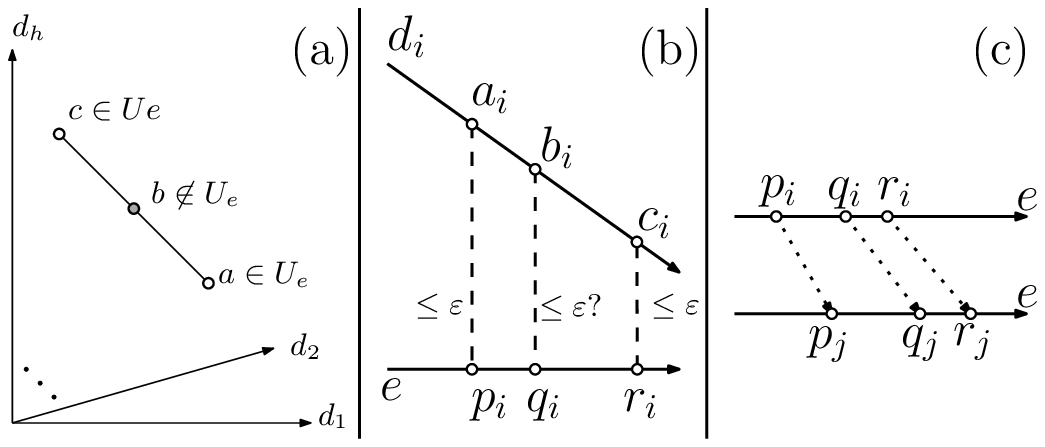}
     \centering \caption{\small (a) $a,c \in U_e$ but $b \not\in U_e$. (b) both $a_i$ and $c_i$ being within distance $\varepsilon$ of their
respective points on $e$ necessitates that $b_i$ is within distance $\varepsilon$ of its own. (c) The crossing points of diagonals $d_i$ and $d_j$ being in order along $e$ for both points $a$ and $c$ necessitates that they are also in order for $b$}
     \label{fig:UntangleSpace-03}
\end{figure}

\proof
To construct a proof we first define $U_{e}$ more formally.

\begin{eqnarray*}
U_e(d_1, \ldots, d_h) = \lbrace a=(a_1,\ldots,a_h) | (\exists p_1, \ldots, p_h \in e | (\forall i,j |  (a_i \in d_i, 1 \leq i \leq j \leq h) \rightarrow \\ ( \|a_i-p_i\| \leq \varepsilon, p_i < p_j ))) \rbrace
\end{eqnarray*}

where $p_i \leq p_j$ means that the point $p_i$ is no further than the point $p_j$ along the edge $e$.  If $U_{e}$ is not convex then there are points $a=(a_1,\ldots,a_h), c=(c_1,\ldots,c_h) \in U_{e}$ such that there exists a point $b=(b_1,\ldots,b_h)$ on the line segment between $a$ and $c$ where $b \not\in U_{e}$, see Figure \ref{fig:UntangleSpace-03}a.  The point $b$ is on the line segment between $a$ and $c$ so there exists an x between $0$ and $1$ such that $b = xa + (1-x)c$.  Likewise for each point $b_i \in b$, $b_i = xa_i + (1-x)c_i$.  Let $p_i$ and $r_i$ be the points on edge $e$ mapped to by pointa $a_i$ and $b_i$ respectively.  Let $q_i = xp_i + (1-x)r_i$.  Since $a,c \in U_{e}$, $\|a_i-p_i\| \leq \varepsilon$ and $\|c_i-r_i\| \leq \varepsilon$.  It follows from these that $\|b_i-q_i\| \leq \varepsilon$, see Figure \ref{fig:UntangleSpace-03}b.

\begin{eqnarray*}
\|a_i-p_i\| \leq \varepsilon &\Rightarrow& x\|a_i-p_i\| \leq x\varepsilon \\
                             &\Rightarrow& \|xa_i-xp_i\| \leq x\varepsilon \\
& & \\
\|c_i-r_i\| \leq \varepsilon &\Rightarrow& (1-x)\|c_i-r_i\| \leq (1-x)\varepsilon \\
                             &\Rightarrow& \|(1-x)c_i-(1-x)r_i\| \leq (1-x)\varepsilon \\
                             &\Rightarrow& \|(1-x)c_i-(1-x)r_i\| \|xa_i-xp_i\| \leq (1-x)\varepsilon + x\varepsilon \\
                             &\Rightarrow& \|(xa_i + (1-x)c_i)-(xp_i + (1-x)r_i)\| \leq \varepsilon - x\varepsilon + x\varepsilon \\
                             &\Rightarrow& \|b_i-q_i\| \leq \varepsilon
\end{eqnarray*}

Since $a,c \in U_{e}$, $p_i \leq p_j$ and $r_i \leq r_j$.  It follows from a proof similar to the previous one that $q_i \leq q_j$, see Figure \ref{fig:UntangleSpace-03}c.

\begin{eqnarray*}
p_i \leq p_j &\Rightarrow& xp_i \leq xp_j \\
& & \\
r_i \leq r_j &\Rightarrow& r_i \leq r_j \\
             &\Rightarrow& (1-x)r_i \leq (1-x)r_j \\
             &\Rightarrow& xp_i + (1-x)r_i \leq xp_j + (1-x)r_j \\
             &\Rightarrow& q_i \leq q_j
\end{eqnarray*}

Thus, for all $i,j$, where $0 \leq i \leq j \leq h$, $\|b_i-q_i\| \leq \varepsilon$ and $q_i \leq q_j$.  From this it follows by definition that $b \in U_{e}$.  Therefore $U_{e}$ is convex.\qed

\section{Constant Factor Approximation Order of Regions \label{sec:CFATechnicalCases}}

We now need to show that the points that the intersections are mapped to do not occur out of order.  This could happen when the preimage regions on $d$ occur out of order, see Figure \ref{fig:FoldedApprox-02}(c).  In this case, collapsing to the leftmost point of a region for a single intersection would not be sufficient.  The two collapsed points would still be out of order along $d$.

We consider two cases.  For each of these cases we assume that the intersection points $a$ and $b$ are at on the top level of the arrangement of the image curves.

\begin{figure}[ht]
     \centering \includegraphics[width=0.95\textwidth]{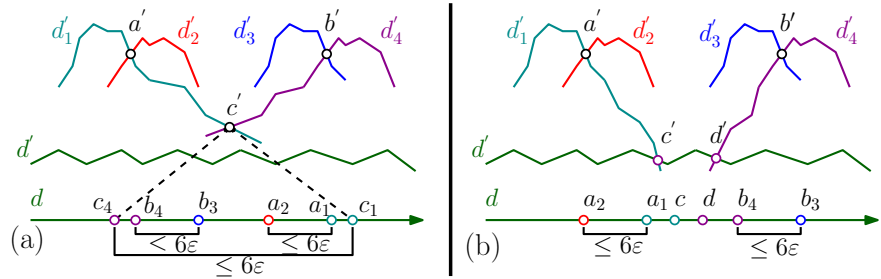}
     \centering \caption{\small
         (a) If the image curves intersect before crossing $d^{'}$, the preimages of both intersections must completely occur on a region of 
             $d$ smaller than $6\varepsilon$,
         (b) If the image curves intersect $d^{'}$ before each other, the preimages of the intersections must occur in the proper order}
     \label{fig:FoldedApprox-06}
\end{figure}

\subsection*{Case I\label{sec:GeneralApproxCaseI}}

See figure Figure \ref{fig:FoldedApprox-06}(a).  Consider the image curve of the first intersection which is sloping toward $d^{'}$ ($d_1^{'}$ in example) and the image curve of the second intersection which is sloping away from $d^{'}$ ($d_4^{'}$ in example) intersect each other before intersecting $d^{'}$ (moving left-to-right).  Assume $d_1^{'}$ first intersects $d_2^{'}$ and then $d_4^{'}$, and assume $d_4^{'}$ first intersects $d_2^{'}$ and then $d_3^{'}$.

In this case we still have a correspondence between the two diagonals and $d$ (they are still above d and thus a mapping exists).  The preimages of the intersection points on $d$ are restricted to being outside the regions from Figure \ref{fig:FoldedApprox-02}(c).  The region they enclose must be $\leq 6\varepsilon$ in width.  Thus we can just map the whole region to the leftmost point.

\subsection*{Case II\label{sec:GeneralApproxCaseII}}

See Figure \ref{fig:FoldedApprox-06}(b).  The image curves cross $d^{'}$ before they intersect each other.  The order the curves intersect in places a restriction on how far back the preimage of intersection 2 can be.  Specifically, in this case, the preimages of the two intersections must occur in order on $d$ so there is no new restriction imposed by them.

\subsection*{Induced Mapping\label{sec:InducedMapping}}

One of these two cases must occur.  In particular, the case where the image curves cross earlier or later than specified in Case I cannot happen.  In this hypothetical case one of the image curves would be forced above the other one.  Assume $d_1^{'}$ is the one forced above.  In that case $d_1^{'}$ would need to pass above the intersection $b^{'}$ in order to not cross $d_4^{'}$.  We assumed earlier that $b^{'}$ was at the highest level of the arrangement so this is a contradiction.  The same argument works if $d_4^{'}$ is the image curve forced above.

From these two cases we can see that monotonicity constraints will not occur between different intersection preimages on $d$.  Thus, we can collapse each of them to a single point using the $9\varepsilon$ approximation described above without the collapsed points occurring out of order.

\section{Remaining Cases for Lemma \ref{lem:HalfSpaceRestrictions} \label{sec:Lemma4Appendix}}

\begin{figure}[ht]
     \centering \includegraphics[width=0.95\textwidth]{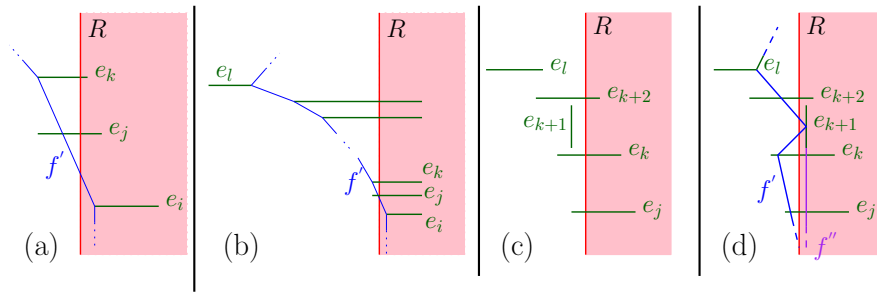}
\centering \caption{\small  (a) the edge $e_{k+1}$ forces $f^{'}$ out of $R$ on $e_j$, but for this to happen $e_{k+1}$ must be completely outside of $R$, thus no $f$ can cross $e_{k+1}$.  (b) $f^{'}$ is not forced out of $R$ on $e_j$ but rather can follow the path $f^{''}$.  (c) the path can continue to curve away from $R$. (d) in this case, while the shortest path goes out of $R$ eventually there must be an edge $e_l$ where it reverses direction since $f^{'}$ ends at $b$ which is in $R$.  Such an edge cooresponds to case (I).}
     \label{fig:ParallelPerpTangleExpNew-03}
\end{figure}

Case (III) occurs when part of $e_k$ is in $R$ and $f^{'}$ crosses it in the part outside of $R$, see Figure \ref{fig:ParallelPerpTangleExpNew-03}(a).  The path $f$ can cross the part of $e_k$ inside $R$.  This case can be repeated many times but eventually, $f^{'}$ ends at $b$ so it needs to cross back into $R$.  To do this it must cross an edge on the opposite end point.  If that end point is in $R$ the entire path can be shortcut similar to case (II) and if the point is not in $R$ then as in case (I) there does not exist an $f$ which can cross $e_l$ and remain in $R$ see Figure \ref{fig:ParallelPerpTangleExpNew-03}(b).  Thus case (III) also leads to a contradiction.

Next we consider the two cases that arise from adding a perpendicular fold $e_{k+1}$.  Specifically, we will consider where the first perpendicular fold after $e_j$ is placed.  Such a fold runs parallel with $\partial R$ so it must be either entirely inside $R$ or entirely outside.

Case (IV) occurs if the edge $e_{k+1}$ is outside of $R$.  In this case there does not exist an $f$ which can pass through $e_{k+1}$ and be entirely in $R$, see Figure \ref{fig:ParallelPerpTangleExpNew-03}(c).

Case (V) occurs if the fold $e_{k+1}$ is inside of $R$.  In this case the shortest path $f^{'}$ will no longer be forced outside of $R$ on the edge $e_{j+1}$.  Instead, it goes through $R$ to the edge $e_{k+1}$, see Figure \ref{fig:ParallelPerpTangleExpNew-03}(d).  The only way that $f^{'}$ could not pass through $R$ would be if one of the edges between $e_{j}$ and $e_{k}$ were entirely outside $R$ and that is ruled out by case (I) 

Thus no such $Q$, $R$, and $f$ can exist.

\end{document}